\def\beq{\begin{equation}}
\def\eeq{\end{equation}}
\def\tday{t_{\rm day}}
\documentstyle[12pt,aasms4]{article}
\begin{document}

\title{A Possible Explanation for the Radio ``Flare''
in the Early Afterglow of GRB990123}

\author{Xiangdong Shi and Geza Gyuk}
\affil{Department of Physics, University of California, San Diego,
La Jolla, CA 92093}
\authoremail{shi@physics.ucsd.edu, gyuk@mizar.ucsd.edu}

\begin{abstract}
We suggest that the deceleration of the relativistic shock by
a denser part of the interstellar medium off line-of-sight produced
the observed radio ``flare'' in the early afterglow of GRB990123.
We find that this scenario is consistent with observations if the
particle number density of this denser part of the medium is between
$\sim 200$ and $\sim 2\times 10^4$ cm$^{-3}$.  Because of the
premature deceleration of part of the shock, the later stage of
the afterglow should decay modestly faster than the powerlaw 
expected from an isotropic shock propagation.
\end{abstract}


\section{Introduction}
GRB990123 had some remarkable characteristics that drew intense interest
from astrophysicists.  Its source appears to be at a cosmological redshift
$z\ge 1.61$ (Kelson et al. 1999).  Assuming isotropic emission this leads
to a huge energy release: $\ga 3\times 10^{54}$ ergs in $\gamma$-rays
alone (Kulkarni et al. 1999a). Prompt optical follow-ups saw a very bright
(magnitude 9) early afterglow optical flash (Akerlof et al. 1999). Since
the $\gamma$-ray burst (GRB) itself, its afterglow has been monitored
almost constantly in X-rays, the optical band and radio.

Radio observations of GRB990123 have been particularly puzzling. As
expected from afterglow models, initial measurements (six hours after the
burst) obtained only upper limits (Frail et al. 1999a). One day later,
however, new observations showed a 8-$\sigma$ detection at 8.46 GHz of
260$\pm 32 \mu$Jy (Frail et al. 1999b).  On the other hand, observations
at 4.88 GHz made during a period overlapping with the 8.46 GHz detection,
yielded only upper limits (Galama et al. 1999). Finally, three to five
days after the burst the afterglow radio output was again consistent with
zero (Kulkarni et al. 1999b). Figure 1 summarizes the detection and upper
limits.  The radio emission is rather peculiar, as it does not conform to
the gradual rise-and-decay time profile with a timescale of ${\cal O}(10)$
days which is expected from the standard GRB afterglow model and which was
rather successfully confirmed by the radio afterglow observations of
GRB970508 (Waxman, Kulkarni and Frail 1998).

Several possibility have been raised to explain this one-time radio
``flare''. Kulkarni {\sl et al.} (1999b) have suggested that the
flare is an interstellar scattering and scintillation
event. Alternatively, it might be part of the early afterglow from the
external reverse shock, if self-absorption of the radio emission
suppressed the radio flux during the first day as well as during the
second day but only at the 4.88 GHz band (Sari and Piran 1999).

Here we investigate a third possibility (Shi and Gyuk 1999),
that the radio ``flare'' might be due to the relativistic
shock ploughing into a dense part (a cloud, or ejecta, for example)
of the interstellar medium (ISM) off line-of-sight (LOS).  This off LOS
portion of the shock was therefore decelerated more efficiently than the
rest (including that along LOS), and in turn gave rise to the premature
emission of radio which also faded away relatively rapidly.
We envision a geometry as in Figure 2.  We will refer generally
to this denser part of ISM off the LOS as a ``cloud''.

In the standard GRB afterglow model, the afterglow is generated by
synchrotron radiation in the external shock of the GRB event as the shock
gradually decelerates in a homogeneous ISM (with density $n\sim 1$
cm$^{-3}$).  The frequency of the synchrotron radiation depends strongly
on the Lorentz factor, $\gamma_e$, of the electrons in the shock, which in
turn scales with the Lorentz factor, $\gamma$, of the shock. Therefore,
the afterglow starts at shorter wavelengths, in X-rays, and progressively
moves to longer wavelengths, as the shock gradually slows down in the
ISM. After $\cal O$(10) days, the afterglow peaks in the radio band. A
similar progression is envisioned to occur in the portion of the shock
that encounters the off-LOS cloud, except over a much shorter
timescale. However, because the afterglow radiation is beamed to an
opening angle $\sim 1/\gamma$ we can only see the off-LOS cloud shock if
it is within $\sim 1/\gamma$ of the LOS.  Thus relativistic beaming
alone may prevent short wavelength (optical etc.) ``flares'' originating
from the off-LOS cloud from being detectable.

If the size of the cloud is comparable to the distance of the cloud to the
GRB source, we can crudely approximate the deceleration of the
relativistic shock in the cloud as if the shock were decelerating in an
homogeneous ISM of enhanced particle density ($n\gg 1$ cm$^{-3}$) and with
spherical symmetry.  This approximation should hold sufficiently well for
the later epoch of the deceleration, which is relevant to radio emission.
In so doing, we employ the scaling relations in the standard
afterglow model.

The Lorentz factor of the relativistic shock scales as
\beq	
\gamma\approx\left\{\begin{array}{ll}
 6\,E_{52}^{1/7}n_1^{-1/7}\gamma_{100}^{-1/7}\tday^{-3/7}[(1+z)/2.6]^{3/7},
 &\mbox{for radiative shocks;}\\
 7\,E_{52}^{1/8}n_1^{-1/8}\tday^{-3/8}[(1+z)/2.6]^{3/8},
 &\mbox{for adiabatic shocks.}\end{array}\right.
\label{gammascaling}
\eeq
from energy and momentum conservation considerations (see e.g., Piran
1998). In equation (1), $E_{52}$ is the initial energy of the shock in
units of $10^{52}$ erg, assuming a $4\pi$ expansion angle; $n_1$ is the
particle number density of the medium in cm$^{-3}$; $\gamma_{100}$ is the
initial Lorentz factor of the shock in units of 100; $\tday$ is the time
elapsed since the GRB in days, as seen by the observer; and $z$ is the
redshift of the GRB.  In the radiative regime, the particles in the shock
convert their kinetic energy into radiation rather efficiently. In the
adiabatic regime, the radiation loss is negligible.

For the external shock in a GRB, the transition from the radiative
regime to the adiabatic regime occurs at a time (Piran 1998)
\begin{equation}
t_{\rm r\rightarrow a}\sim 0.002\,E_{52}^{4/5}n_1^{3/5}(\epsilon_e/0.6)^{7/5}
(\epsilon_B/0.01)^{7/5}[(1+z)/2.6]^{12/5}\gamma_{100}^{-4/5}\,{\rm day},
\label{r-a transition}
\end{equation}
where $\epsilon_e$ is the fraction of thermal energy of the shock that
resides in the random motion of electrons, and $\epsilon_B$ is the
ratio of the magnetic field energy to the thermal energy density
of the shock ($\sim 4\gamma^2n_1m_pc^2$ where $m_p$ is the proton
mass and $c$ the speed of light).
Canonical values are $\epsilon_e\sim 0.6$ and $\epsilon_B\sim 0.01$,
obtained by fitting the standard afterglow model to the observed afterglow
of GRB970508 (Wijers and Galama 1998; Granot, Piran and Sari 1998).
Before $t_{\rm r \rightarrow a}$ the cooling time is shorter than the
dynamic timescale, and the shock is radiative.

For an energetic GRB event such as GRB990123 where $E_{52}\sim 10^3$, and
a cloud much denser than the average ISM ($n_1\gg 1$), the transition
occurs much later than a day. We therefore assume a radiative shock in our
treatment.

There are three synchrotron emission frequencies that are crucial:
$\nu_m$, the peak synchrotron radiation frequency if electrons in
the shock are slow-cooling; $\nu_c$, the peak synchrotron radiation
frequency if electrons are fast-cooling; and $\nu_a$, the
self-absorption frequency of the synchrotron radiation below
which the radiation is absorbed by electrons in the shocks.
Depending on the ratio of these key frequencies, the synchrotron
radiation from the external shock can have very different
spectral shapes.

In the standard afterglow picture, the shock-heated electrons
develop a power law number density distribution $N(\gamma_e)
\propto\gamma_e^{-p}$ where $\gamma_e\ge\gamma_{e,min}$ is
the Lorentz factor of electrons. The minimum Lorentz factor
cut-off is
\begin{equation}
\gamma_{e,min}\approx {p-2\over p-1}{m_p\over m_e}\epsilon_e\gamma
\sim 2.2\times 10^3\,(\epsilon_e/0.6)E_{52}^{1/7}n_1^{-1/7}
\gamma_{100}^{-1/7}\tday^{-3/7}[(1+z)/2.6]^{3/7},
\label{slowcool}
\end{equation}
where $m_p$ and $m_e$ are proton and electron masses respectively
(Piran 1998).
The power law index $p$ is found to be $\sim 2.5$ by fitting the
observed GRB spectra and that of the afterglows.  If electrons are
slow-cooling, their peak synchrotron emission will be in the
observer's frame at a frequency
\beq
\nu_m ={\gamma\gamma_{e,min}^2\over 1+z}{eB\over 2\pi m_ec}\approx 
7\times 10^{12}\,(\epsilon_B/0.01)^{1/2}(\epsilon_e/0.6)^2E_{52}^{4/7}
n_1^{-1/14}\gamma_{100}^{-4/7}\tday^{-12/7}[(1+z)/2.6]^{5/7}\,{\rm Hz},
\label{nu_m}
\eeq
where $e$ is the electron charge and $B$ is the magnetic field.

If, however, the shocked electrons cool quickly, they will mostly
radiate from a cooled state, whose Lorentz factor is
\beq
\gamma_{e,c}\sim {3m_ec\over 4\sigma_T(B^2/8\pi)\gamma t}
\approx 7\times 10^4\,(\epsilon_B/0.01)^{-1}
E_{52}^{-3/7}n_1^{-4/7}\gamma_{100}^{3/7}
\,\tday^{2/7}\,[(1+z)/2.6]^{-2/7}
\label{fastcool}
\eeq
where $\sigma_T=8\pi e^4/3m_e^2c^4=6.65\times 10^{-25}$ cm$^2$
is the Thompson scattering cross section (Piran 1998). The
emitting frequency in the observer's frame is
\begin{equation}
\nu_c ={\gamma\gamma_{e,c}^2\over 1+z}{eB\over 2\pi m_ec}
\approx 5\times 10^{15}\,(\epsilon_B/0.01)^{-1.5}\,E_{52}^{-4/7}\,
n_1^{-13/14}\,\gamma_{100}^{4/7}\,\tday^{-2/7}\,[(1+z)/2.6]^{-5/7}\,{\rm Hz}.
\label{nu_c}
\end{equation}

To find the self-absorption frequency $\nu_a$, we follow Granot {\sl et
al.}  (1999) and Wijers and Galama (1999) to calculate at what frequency
the optical depth becomes unity.  A crude estimate of the optical depth
$\tau$ is $\tau\sim \alpha^\prime_{\nu^\prime} R/\gamma$, where
$\alpha^\prime_{\nu^\prime}$ is the absorption coefficient at a
a frequency $\nu^\prime$, and $R/\gamma$
is the thickness of the shock, all in the local rest frame.  However, we
cannot directly adopt the formula for $\nu_a$ in Granot {\sl et al.}
(1999), or Wijers and Galama (1999), because both have assumed a slow
electron cooling regime in an adiabatic shock.  In our problem, the
electrons are fast-cooling, and the shock is radiative.  We therefore
substitute the electron Lorentz factor $\gamma_{e,c}$ (fast-cooling
regime) for $\gamma_{e,min}$ (slow-cooling regime), and likewise
substitute the shock Lorentz factor $\gamma$ in radiative shocks for that
in adiabatic shocks.  The resultant self-absorption frequency is then
\begin{equation}
\nu_a\sim 10^8\,(\epsilon_B/0.01)^{6/5}\,E_{52}^{4/5}\,n_1\,\gamma_{100}
^{4/5}\,\tday^{-4/5}\,[(1+z)/2.6]^{-1/5}\,{\rm Hz}.
\label{nu_a}
\end{equation}
We have implicitly assumed $\nu_a\la\nu_c$ since for rapid cooling 
most of the electrons will be at Lorentz factor $\gamma_{e,c}$.
For these electrons, absorption at frequencies $\gg\nu_c$ falls off
rapidly.  If we further assume that the time-integrated absorption
from newly injected electrons in transition to their final cooled 
state is small, we will always have $\nu_a\la\nu_c$.

Therefore, with $E_{52}\sim 10^3$, $n_1\gg 1$ and $\tday\sim 1$,
and canonical values of $\epsilon_B$, $\epsilon_e$ and $\gamma_{100}$,
there is a hierarchy in frequencies: $\nu_a\la \nu_c < \nu_m$.
The peak flux of the synchrotron radiation seen by an observer
is at $\nu_c$ (Piran 1998):
\beq
F_{\nu}(\nu_c)=F_{\nu,max}\approx 1.8\times 10^3\,{\cal F}
(\epsilon_B/0.01)^{1/2}\,E_{52}^{8/7}\,n_1^{5/14}\,\gamma_{100}^{-8/7}\,
\tday^{-3/7}\,[(1+z)/2.6]^{10/7}\,d_{28}^{-2}\,\mu{\rm Jy},
\label{Fnumax}
\eeq
where ${\cal F}<1$ is a geometric factor to account for the fact that
the emission is from off LOS so we only see an edge of the radiation
cone, and $d_{28}$ is the luminosity distance in the units of 10$^{28}$
cm. Fluxes at other frequencies are (Piran 1998)
\beq
F_{nu}\approx\left\{\begin{array}{ll}
 (\nu/\nu_a)^2F_\nu(\nu_a) & \mbox{if $\nu<\nu_a$;} \\
 (\nu/\nu_c)^{1/3}F_\nu(\nu_c) & \mbox{if $\nu_a<\nu<\nu_c$;} \\
 (\nu/\nu_a)^{-1/2}F_\nu(\nu_c) & \mbox{if $\nu_c<\nu<\nu_m$;} \\
 (\nu/\nu_a)^{-p/2}F_\nu(\nu_m) & \mbox{if $\nu>\nu_m$.} 
\end{array}\right.
\label{spectrum}
\eeq
The spectrum and its evolution is schematically plotted in Figure 3.

We require $\nu_c,\nu_a>8.46$ GHz at the time of $\tday\sim 1$
so that a non-detection of radio signal at 4.88 GHz is compatible
with a simultaneous detection at 8.46 GHz. This condition is 
satisfied if $n_1\la 2\times 10^4$. Assuming at a redshift 
$z\approx 1.61$, the luminosity distance to GRB990123 is
$d_{28}\sim 4$ (its order of magnitude is insensitive to
different choices of cosmology). The peak flux is then
$F_{\nu,max}\sim 3\times 10^5\,{\cal F}\,n_1^{5/14}\,\mu$Jy
when adopting for other parameters values mentioned above. 
Scaling down to 8.46 GHz, we find $F_\nu(8.46\,{\rm GHz})
\sim F_{\nu,max}(8.46\,{\rm GHz}/\nu_c)^2 \sim 3\times 10^{-3}
\,{\cal F}\,n_1^{31/14}\,\mu$Jy.\footnote{We have assumed 
$\nu_a\sim\nu_c$, which will be the case for $n_1\ga 200$.}
To yield a 260 $\mu$Jy detection would therefore require
$n_1\ga 200$. Because at this part of the spectrum $F_{\nu}\propto 
\nu^2$, a flux of 260$\pm 32\,\mu$Jy at 8.46 GHz implies a flux of
86$\,\mu$Jy at 4.88 GHz, consistent with the $3\sigma$ limit of 
130$\,\mu$Jy measured at this frequency (Galama et al. 1999). 
Given $200\la n_1\la 2\times 10^4$,  the Lorentz factor of the
prematurely decelerated portion of the shock at $\tday\sim 1$ 
is of order ${\cal O}(1)$. 

The non-detection of radio emission six hours after the burst may be due
to strong absorption (i.e., $\nu_a$ too large), or it may simply be that
the shock hadn't yet encountered the cloud.  While three days later, this
portion of the shock has become very weak, and its emission is further
absorbed by the main shock that propagates along LOS. It should be kept in
mind that a factor of several below the level of the detected emission
might render the emission undetectable.

Assuming that the dimension of the cloud is comparable to the size
of the fireball $\sim 4\gamma^2\,t$ at $\tday\sim 1$, we find a mass
for the cloud to be $\sim 10^{-5}$ to $10^{-3}\,M_\odot$. We speculate
that it may be ejecta from the GRB site.

The main portion of the relativistic shock along LOS is not affected by
the cloud off LOS. It generates the main afterglow as expected from the
standard GRB afterglow model.  The temporal decay of this afterglow in a
given frequency band follows a powerlaw $t^{-1.1}$ (Vietri 1997; Waxman
1997; Sari, Piran and Narayan 1998).  But as its radiation cone become
wider (opening angle $\theta\sim 1/\gamma$), the viewing area of an
observer is larger.  Eventually the area will engulf the portion of the
shock that was prematurely decelerated and terminated.  The temporal decay
of the main afterglow should then be faster than $t^{-1.1}$.  The
transition to a faster decay law is not unique: for example, if the
relativistic shock is a narrow jet, the temporal decay of its afterglow
steepens when its opening angle $\theta < 1/\gamma$. Depending on the
details of the model it may either steepen by an additional $t^{-3/4}$
power (M\'esz\'aros \& Rees 1999), or steepen to $t^{-p}$ (Kulkarni et
al. 1999). The rate of afterglow decay due to an off-LOS hole in a
spherical shock should be more modest than that due to a jet and indeed in
this scenario, we should expect the afterglow decay will eventually
approaches its initial shallower decay profile, as
the influence of the geometric defect becomes increasingly less
significant.

In summary, we show that the radio ``flare'' observed in the early
afterglow of GRB990123 may be due to a relativistic shock 
encountering a denser part of the ISM, (with a density between
$\sim 200$ and $\sim 2\times 10^4$ cm$^{-3}$)
off line-of-sight. A transition from a $t^{-1.1}$ decay to a
modestly faster temporal decay is expected in the later stage
of the afterglow. This scenario also implies that the relativistic
shock that generates the afterglow is unlikely to be beamed
by more than a factor of a few. 

We thank George Fuller, Bob Gould and Art Wolfe for discussions.  XS
acknowledges support from NSF grant PHY98-00980 at UCSD.  GG wishes to
thank the Department of Energy for partial support under grant
DEFG0390ER40546 and Research Corporation.

\newpage
\begin{figure}[tbp]
\begin{center}
\epsscale{1.} \plotone{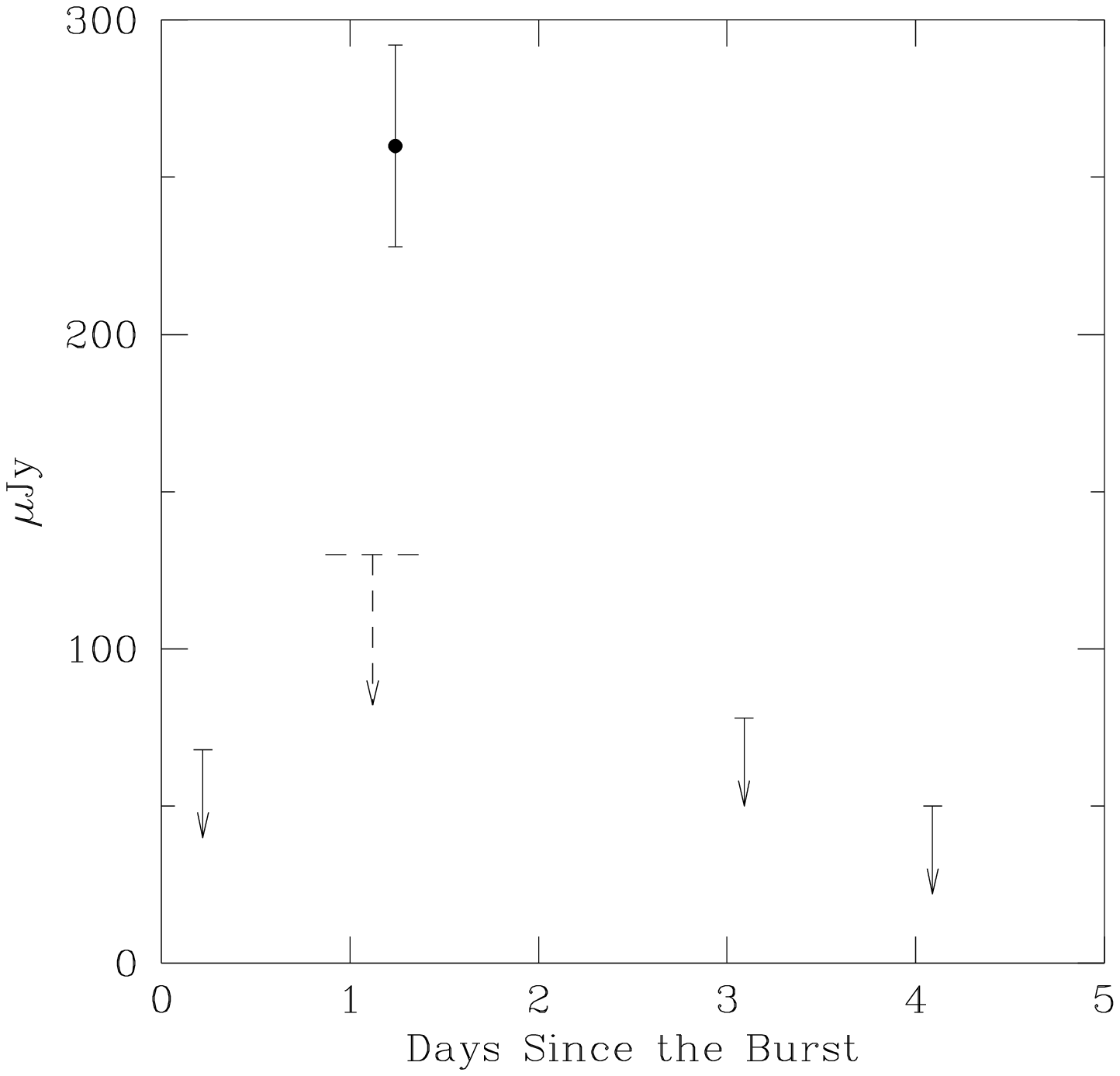}
\end{center}
\caption{Radio observations at 8.46 GHz (the solid lines, upper limits
are $2\sigma$, Kulkarni et al. 1999b) and 4.88 GHz (the dashed lines,
$3\sigma$ upper limits, Galama et al. 1999).}
\label{fig1}
\end{figure}

\newpage
\begin{figure}[tbp]
\begin{center}
\epsscale{1.} \plotone{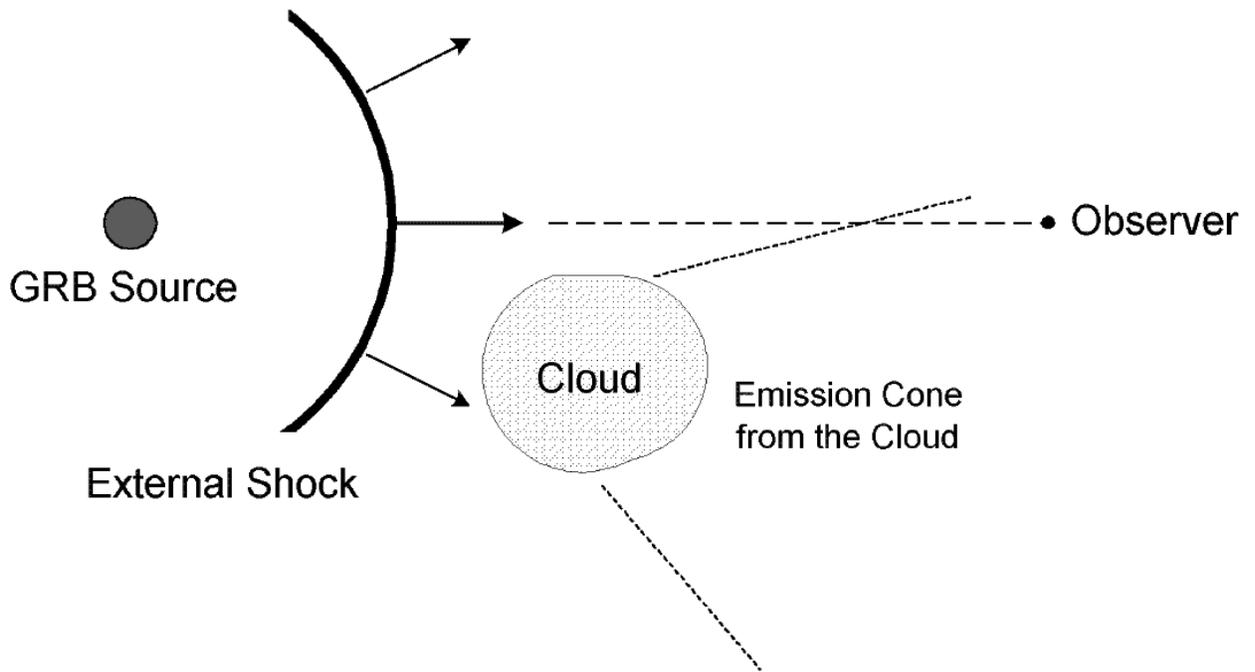}
\end{center}
\caption{A possible geometry near the site of GRB990123.}
\label{fig2}
\end{figure}

\newpage
\begin{figure}[tbp]
\begin{center}
\epsscale{1.} \plotone{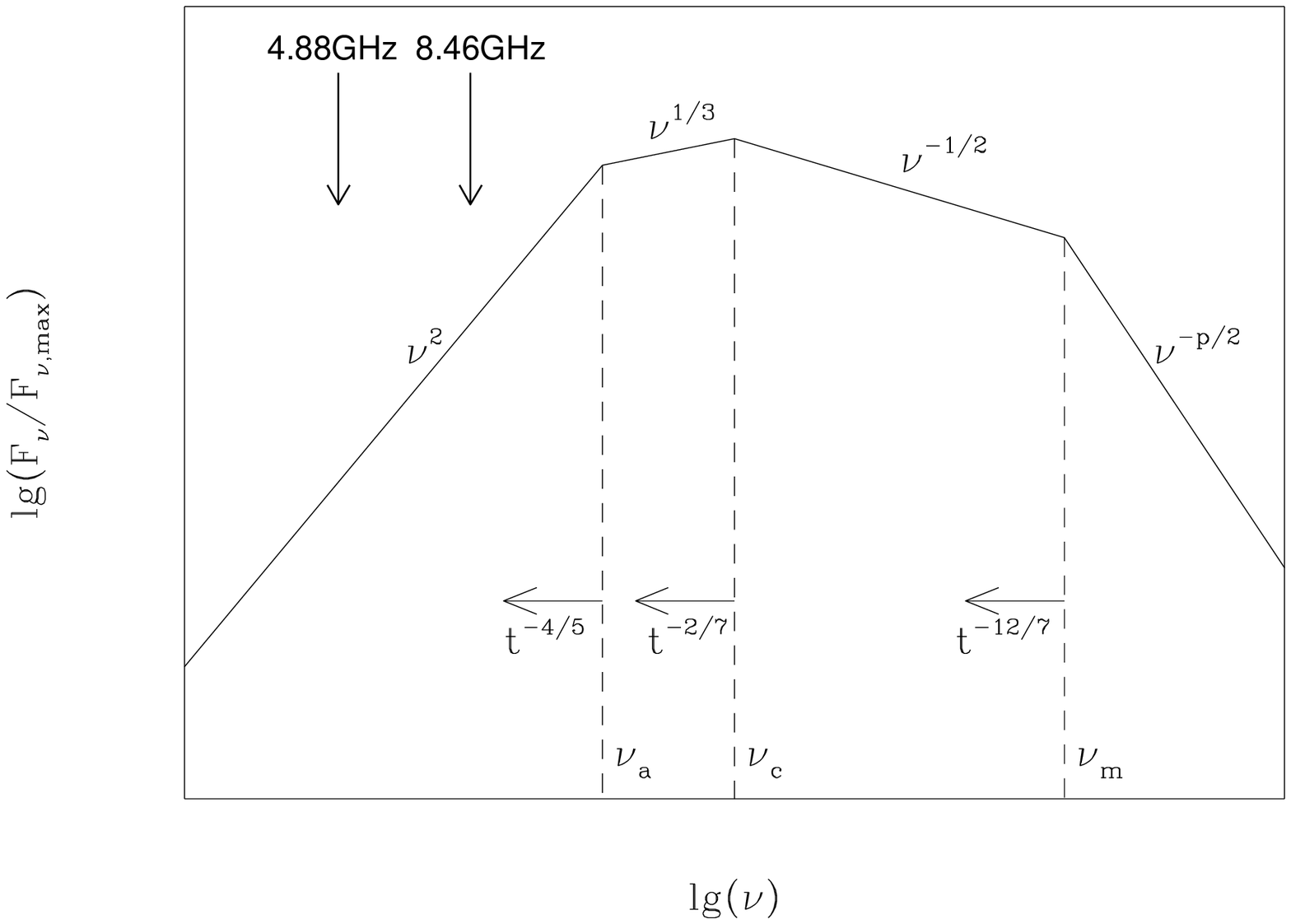}
\end{center}
\caption{A schematic plot of the afterglow spectrum from the off
line-of-site deceleration in the cloud.}
\label{fig3}
\end{figure}


\begin{thebibliography}{}
\bibitem{Akerlof} Akerlof, C. et al. 1999, GCN 205
\bibitem{Frail} Frail, D.~A. et al. 1999a, GCN 200
\bibitem{Frail} Frail, D.~A. et al. 1999b, GCN 211
\bibitem{Galama} Galama, T.~J. et al. 1999, GCN 212
\bibitem{Granot} Granot, J., Piran, T. and Sari, R. astro-ph/9808007
		 (unpublished)
\bibitem{Kelson} Kelson, D.~D., Illingworth, G.~D., Franx, M. Magee, D.,
		 van Dokkum, P.~G. 1999, {\sl I.A.U.C.} 7096
\bibitem{Kulkarni} Kulkarni, S.~R. et al., 1999a, submitted to Nature
\bibitem{Kulkarni} Kulkarni, S.~R. et al. 1999b, GCN 239
\bibitem{Meszaros} M\'esz\'aros, P., and Rees, M.~J. 1999, 
		   astro-ph/9902367 (unpublished)
\bibitem{Piran} Piran, T. 1998, to appear in Physics Reports
\bibitem{Rybicki} Rybicki, G.~B. and Lightman, A.~P. 1979, Radiative
		  Processes in Astrophysics (New York: John Wiley \& Sons)
\bibitem{Sari0} Sari, R., Piran, T. and Narayan, R. 1998, ApJ, 497, L17
\bibitem{Sari} Sari, R. and Piran, T. 1999, astro-ph/9802009 (unpublished) 
\bibitem{Shi} Shi, X. and Gyuk, G. 1999, GCN 247
\bibitem{Vietri} Vietri, M. 1997, ApJ, 488, L105
\bibitem{Waxman0} Waxman, E. 1997, ApJ, 485, L5
\bibitem{Waxman} Waxman, E., Kulkarni, S.~R. and Frail, D.~A. 1998, ApJ, 497,
		 288
\bibitem{Wijers} Wijers, R.~A.~M.~J. and Galama, T.~J. 1998, astro-ph/9805341
		 (unpublished)
\end{thebibliography}
\end{document}